\documentclass[aps,twocolumn,pra,floatfix]{revtex4-1}

\usepackage{dcolumn}
\usepackage{amssymb}
\usepackage{amsmath}
\usepackage{graphicx}
\usepackage{nicefrac}

\include{phys_jdf}

\newcommand{\E}[1]{\ensuremath{\times 10^{#1}}}

\newcommand{\Wtarget}{\ensuremath{\textrm{W}^{20+}}}
\newcommand{\Wcompound}{\ensuremath{\textrm{W}^{19+}}}
\newcommand{\eps}{\ensuremath{\varepsilon}}
\newcommand{\Jpi}{\ensuremath{J^\pi}}
\newcommand{\GammaSpr}{\ensuremath{\Gamma_{\rm spr}}}

\newcommand{\eref}[1]{(\ref{#1})}
\newcommand{\Fig}[1]{Fig.~\ref{#1}}

\begin{document}

\title{Level-resolved quantum statistical theory of electron capture into many-electron compound resonances in highly charged ions}

\author{J. C. Berengut}
\author{C. Harabati}
\author{V. A. Dzuba}
\author{V. V. Flambaum}
\affiliation{School of Physics, University of New South Wales, Sydney NSW 2052, Australia}
\author{G. F. Gribakin}
\affiliation{School of Mathematics and Physics, Queen's University, Belfast BT7 1NN, Northern Ireland, United Kingdom}

\date{4 November 2015}

\begin{abstract}

The strong mixing of many-electron basis states in excited atoms and ions with open $f$ shells results in very large numbers of complex, chaotic eigenstates that cannot be computed to any degree of accuracy. Describing the processes which involve such states requires the use of a statistical theory. Electron capture into these `compound resonances' leads to electron-ion recombination rates that are orders of magnitude greater than those of direct, radiative recombination, and cannot be described by standard theories of dielectronic recombination. Previous statistical theories considered this as a two-electron capture process which populates a pair of single-particle orbitals, followed by `spreading' of the two-electron states into chaotically mixed eigenstates. This method is similar to a configuration-average approach, as it neglects potentially important effects of spectator electrons and conservation of total angular momentum. In this work we develop a statistical theory which considers electron capture into `doorway' states with definite angular momentum obtained by the configuration interaction method. We apply this approach to electron recombination with W$^{20+}$, considering 2~million doorway states. Despite strong effects from the spectator electrons, we find that the results of the earlier theories largely hold. Finally, we extract the fluorescence yield (the probability of photoemission and hence recombination) by comparison with experiment.

\end{abstract}


\maketitle

\section{Introduction}
Many-body quantum chaos occurs in the excited states of all medium and heavy nuclei \cite{bohr69,zelevinsky96pr}, e.g., the states formed by neutron capture, known as compound resonances. It is also typical in atoms and ions with open $f$ shells. In particular, their excitation spectra demonstrate characteristic Wigner-Dyson level spacing statistics, and the statistics of electromagnetic transition amplitudes is close to Gaussian; both are signatures of quantum chaos \cite{flambaum94pra,gribakin99ajp,flambaum02pra,gribakin03jpb}.
It has been shown that the chaotic mixing of many-electron excited configuration states in such atoms and ions leads to eigenstates that cannot be described using an `exact' theory~\cite{flambaum94pra}. Precise description of these chaotic eigenstates is impossible even in principle, since any minor perturbation (e.g., higher-order correlation corrections or relativistic effects, or neglected interaction with the environment) would lead to radically different mixing of the basis states, due to exponentially small level spacings. In such cases a theory expressed in terms of the exact eigenstates of the Hamiltonian cannot be applied.

Instead, the properties of these systems can be described using a statistical theory of finite quantum systems. This is analogous to classical statistical theories such as the thermodynamics of a hot gas: while the motion of any individual particle cannot be known, the bulk properties such as temperature and pressure can readily be understood by averaging over the many microscopic states corresponding to a small energy interval. Similarly, without an exact description of the compound resonances that describe our system, we can nevertheless calculate properties such as ionization, recombination, and scattering cross sections, using a quantum statistical theory that averages over a small energy interval containing many resonances.
The energy spacing between the compound resonances is usually very small 
(in our example ion \Wcompound\ it is $< 10^{-6}$\,eV) so in experiments this averaging appears naturally.
Earlier papers \cite{flambaum93pst,flambaum93prl} and reviews \cite{flambaum95ppnp,flambaum00pmb}
present the development of the statistical theory for the matrix elements between chaotic compound states. This theory enables one to calculate mean values of orbital occupation numbers, squared electromagnetic
amplitudes, electronic and electromagnetic widths and enhancement of weak interactions in chaotic excited states of  nuclei, atoms and multicharged ions \cite{flambaum94pra,gribakin99ajp,flambaum02pra,gribakin03jpb,flambaum93pst,flambaum93prl,flambaum95ppnp,flambaum00pmb,dzuba12pra,sushkov82ufn,flambaum96pre}.

One use of the statistical theory has been to understand the properties of ions with open $f$-shells. We will consider one of these processes in detail: the recombination of \Wtarget\ with an incident electron to form \Wcompound~\cite{schippers11pra}. Tungsten is a major candidate for the plasma facing components of ITER and future fusion reactors, and processes involving highly-charged tungsten ion are critical for the properties of fusion plasmas \cite{putterich08}.
In the recombination process the incident electron excites one or more target electrons to form a quasistationary resonant state (in simple systems this is a doubly-excited state), which then emits a photon, completing the radiative electron capture \cite{massey42rpp}.
Consideration of this system was motivated by an extant discrepancy, that the measured recombination cross section was much larger than those predicted by very extensive calculations~\cite{badnell12pra}, particularly close to the ionization threshold (i.e., for low incident electron energies). Use of a statistical description of the compound resonances resolved the discrepancy~\cite{dzuba12pra,dzuba13pra}. A similar enhancement of the recombination due to compound resonances was found earlier in the isoelectric ion Au$^{25+}$~\cite{hoffknecht98jpb} and explained by the statistical theory~\cite{flambaum02pra}.

Statistical theory calculations consider the problem as electron capture into ``doorway'' dielectronic states,
which then spread into chaotic compound resonances due to residual
interaction with other valence electrons. This leads to long-time trapping of the incident electron, allowing for the radiative decay to complete the recombination process. Of essential importance is that once the compound resonance has formed, the energy is distributed amongst the many valence electrons, strongly reducing the probability of autoionization and boosting the probability of photoemission and ultimate recombination to be essentially unity (at least near the ionization threshold). Therefore, the problem reduces to calculation of capture into the doorway states, or equivalently, the autoionization rate of the doorways. Further discussion of the role of doorway states in stochastic processes in quantum chaotic systems including atoms, molecules, and nuclei can be found in~\cite{flambaum15pra}.

Earlier statistical theories \cite{flambaum02pra,dzuba12pra,dzuba13pra} treated this problem by combining a many-body theory approach for the
calculation of the amplitude for the two active electrons to populate various single-particle orbitals, with the idea of spreading of the two-electron states into chaotic many-electron eigenstates (i.e., compound resonances). The latter process was parameterized through its rate (known as the spreading width $\Gamma _{\rm spr}$), that was estimated by constructing limited configuration-interaction Hamiltonian matrices. In what follows we refer
to this approach as MBQC (for `many-body quantum chaos') statistical theory.
The above treatment of the dielectronic part of the problem can be described as a `configuration average' approach. We defer details to the next section, but in summary, the doorways into the compound resonances were treated using two-electron wavefunctions, while the remaining electrons (e.g., $4f^7$  in W$^{19+}$) were treated as
`spectators', that formed a spherically-symmetric frozen core, and remained unchanged between the target and doorway states. Our aim is to test this assumption.

In this work we present a `level-resolved' quantum statistical theory in which the doorway wavefunctions include the spectator electrons of the open $f$-shell and are constructed with full account of the total angular momentum of the system. We compare the electron capture cross sections calculated using our level-resolved theory with those of the configuration-averaged MBQC theory. In particular, we examine the effect of the $4f^7$ core on the autoionization rates of the \Wcompound\ doorways. We show that, despite the strong doorway-selectivity enforced by the $4f^7$ electrons, when integrating over all doorways the effect of this core can be neglected in calculation of the capture cross section.

\section{Theory}

In atomic systems that do not exhibit quantum chaotic eigenstates
which involve many active electrons, recombination with a target $A^{q+}$ is usually considered as a sum of direct, radiative recombination (RR) and 
dielectronic recombination (DR). The latter is a two stage process in which an incident electron is captured into a dielectronic resonance of the compound ion, with accompanying promotion of one of the valence electrons in the target,
\begin{equation}
\label{eq:capture}
A^{q+} + e^- \rightarrow A^{(q-1)+**} .
\end{equation}
This is followed by either autoionization, in which case there is no recombination, or radiative relaxation to a level below the ionization threshold, which completes the recombination. The DR process often dominates over the single-electron RR mechanism. Experimentally, much progress has been made due to the use of ion storage rings and electron-beam ion traps (EBITs)~\cite{muller99ptrsl,zou03pra,beiersdorfer08cjp}.
On the theory side, a number of computational approaches have been used successfully to describe DR for many simpler ions and to produce data for plasma modelling (see \cite{nahar94pra,hahn97rpp,lindroth01prl,tokman02pra,badnell03aa,behar04pra,gu08cjp,nikolic09pra,ballance10jpb,badnell11cpc,ballance12jpb} and references therein).

For more complex targets such as Au$^{25+}$, U$^{28+}$, or \Wtarget\ considered in this work, conventional DR calculations underestimate the measured recombination rates, particularly at low incident-electron energy~\cite{badnell12pra,mitnik98pra}. Experiment shows that the recombination rates
at low ($\sim 1$~eV) electron energies in these ions exceed the direct RR rates by two orders of magnitude or more. At the same time the measured rates do not show the sharp resonance structure normally associated with DR \cite{schippers11pra,uwira96hi,hoffknecht98jpb}.

The MBQC statistical theory quantitatively explains the discrepancy as being due to a very dense spectrum of compound resonances: multiply excited, strongly mixed, chaotic many-electron eigenstates.
Note that this situation is distinct from trielectronic recombination (i.e., via resonances with three excited electrons) that has been experimentally observed in Be-like ions (N$^{3+}$, O$^{4+}$, Cl$^{13+}$) \cite{schnell03prl,fogle05aa}. In these systems electron capture into a Rydberg state was accompanied by simultaneous $2s^2\rightarrow 2p^2$ promotions. Trielectronic and quadruelectronic recombination involving resonances with inner-shell excitations was also observed in Li-like to N-like ions of Ar, Fe and Kr \cite{beilmann09pra,beilmann11prl,beilmann13arx}.
However, in the case of chaotic compound resonances one cannot separate out contributions of dielectronic, trielectronic or any other specific resonances with a fixed number of excited electrons. Indeed, a compound state is a chaotic mixture of the states with two, three, four and even five excited electrons, and contributions from all of these configurations are mixed and interfere in the capture amplitude.

Nevertheless, the dielectronic states play a special role.
In the temporal picture, after the capture process \eref{eq:capture}, the dielectronic excitation, which is not an eigenstate of the Hamiltonian, redistributes its energy by populating nearby eigenstates. This can be thought of as a `chain reaction' in which the excited electrons collide with ground-state electrons and excite them. Alternatively, this can be thought of as state mixing in the Hilbert space, leading to ``spreading'' of the initial state and distribution of the expansion coefficients in the exact Hamiltonian basis (i.e., that of the compound resonances). 

After this process has occurred, the initial electron energy is distributed amongst a great number of excited electrons. In a classical picture one can imagine that the probability of any one electron gathering enough energy to overcome the ionization barrier and escape is small. (This is the `trapped billiards' picture used by Bohr to explain the effect of compound resonances in nuclei~\cite{Bohr}.) From a quantum perspective, there are many channels open for radiative decay, and relatively few available for autoionization. In any case, after the internal `spreading' decay of the dielectronic state, the probability of autoionization is strongly suppressed. This is in stark contrast to the standard DR in simple ions where the autoionization rate is typically much larger than the radiative rate.

A more sophisticated treatment of the radiative rate within the quantum statistical theory is provided in \cite{dzuba13pra}, where the fluorescence yield $w_f$ (i.e., the relative probability of photoemission) was calculated for incident electron energies $\eps \leq 120$~eV. It was found that $w_f \sim 1$ for $\eps \rightarrow 0$, but its value quickly drops to around 0.2 for $\eps \gtrsim 15$\,eV. This work is concerned with the electron capture cross section, and we defer complete consideration of the fluorescence yield to a later study. Rather, we will extract the fluorescence yield by comparison of our capture cross section with experiment. The basic equations used to describe resonant recombination are given in the Appendix.
Note that in this work we use atomic units unless otherwise stated, and all energies are taken with respect to the \Wcompound\ ionization threshold, $E_i$.

\subsection{Statistical theory of electron capture into compound resonances}
\label{sec:stat-theory}

Let us write the resonant recombination cross section \eref{cs2} in the following form:
\begin{equation}
\label{cs3}
\sigma= \frac{2\pi^2}{k^2}\sum_{\nu}\frac{(2J_\nu+1)}{2(2J_i+1)}\frac{\Gamma_{\nu i}^{(a)} \Gamma_{\nu}^{(r)}}{\Gamma_\nu}\frac{1}{2\pi}\frac{\Gamma_\nu}
{(\varepsilon-\varepsilon_\nu)^2+\Gamma_\nu^2/4}
\end{equation}
where $i$ indicates the initial target state, the sum is over resonances $\nu$ with energy $\eps_\nu$ (relative to ionization threshold), $\Gamma_{\nu i}^{(a)}$ is the autoionization width $\nu \rightarrow i$ (or equivalently the capture width $i \rightarrow \nu$), $\Gamma_\nu^{(r)}$ is the total radiative decay width, and $\Gamma_\nu$ is the total width of the resonance including all autoionization and radiative decay channels.
If the fluorescence yield
\begin{equation}
\label{bratio}
\omega_f^{(\nu)} = \frac{\Gamma^{(r)}_\nu}{\Gamma_\nu}
\end{equation}
does not change significantly among the resonances, it can be factored out as $\sigma=\omega_f(\varepsilon)\sigma_c$, where
\begin{equation}
\label{ccs0}
\sigma_c= \frac{\pi^2}{k^2}\sum_{\nu}\frac{(2J_\nu+1)}{(2J_i+1)}
\Gamma^{(a)}_{\nu i}
\frac{1}{2\pi}\frac{\Gamma_\nu}{(\varepsilon-\varepsilon_\nu)^2+\Gamma_\nu^2/4}\, 
\end{equation}
is the capture cross section.

When the mixing is strong, each eigenstate 
\begin{equation}\label{eq:eigenket}
|\nu\rangle = \sum_k C_k^{(\nu)} |\varphi_k\rangle 
\end{equation} 
contains a large number $N$ of {\em principal components} $|\varphi_k\rangle $, i.e., basis states for which the expansion coefficients have typical values $C_k^{(\nu)}\sim 1/ \sqrt{N}$ [note the normalization condition Eq.~(\ref{eq:Cnorm})]. The number of principal components can be estimated as $N\sim\Gamma_{\rm spr}/D$, where $D$ is the mean level spacing between the basis states (or eigenstates). Such eigenstates are called compound states. Owing to the strong mixing, the only good quantum numbers that can be used to classify the eigenstates, are the exactly conserved total angular momentum, its projection and parity $J^\pi M $.  The basis set in \eref{eq:eigenket} is formed by constructing linear combinations of Slater determinants, which give eigenstates of $\hat J^2$ and $\hat J_z$ with eigenvalues $J(J+1)$ and $M$, respectively. Such basis states $\varphi_k$ with definite $J$ and $M$ are known as configuration state functions (CSFs). Owing to the chaotic nature of the eigenstates, the capture cross section can be statistically averaged by substituting expansion (\ref{eq:eigenket}) into (\ref{ccs0}), and using the properties of the expansion coefficients, $\overline{C_k^{(\nu)}C_{k'}^{\ast(\nu)}}=\overline {| C_k^{(\nu)}|^2}\delta_{kk'}$.
A full exploration of the statistical properties of the mixing coefficients $C_k^{(\nu)}$ can be found in~\cite{dzuba12pra}.

The energies $E_k=\langle \varphi _k|\hat H|\varphi _k\rangle $ of the principal basis components lie within the spreading width of the eigenenergy $E_\nu$ of the compound state,
$|E_k-E_\nu|\lesssim \Gamma_{\rm spr}$. The components outside the spreading width decrease quickly, so that they do not contribute much to the normalization. For $E_k-E_\nu \approx \mbox{const}$, the components of the chaotic eigenstates have the statistics of Gaussian random variables with zero mean. The variation of their mean-squared value with energy is described well by the Breit-Wigner profile,
\begin{equation}
\label{eq:Csquared}
\overline {\bigl| C_k^{(\nu)}\bigr|^2}=N^{-1}\frac{\Gamma_{\rm spr}^2/4}
{(E_k-E_\nu ) ^2+\Gamma_{\rm spr} ^2/4},
\end{equation}
with $N=\pi \Gamma _{\rm spr}/2D_{J_\nu}$ fixed by normalization
\begin{equation}
\label{eq:Cnorm}
\sum _k\bigl|C_k^{(\nu)}\bigr|^2\simeq
\int \overline {\bigl| C_k^{(\nu)}\bigr|^2} dE_k/D_{J_\nu}=1 \,.
\end{equation}
In fact Eq.~\eref{eq:Csquared} implies that the system is ergodic: all components near a given energy that have the same exact quantum numbers ($J^\pi$, $M$) are mixed with the same average weight.

The sum in Eq.~(\ref{ccs0}) is over the resonances with different $J$, and we can consider the contribution of each $J$ separately. For a fixed $J$,
the sum over the dense spectrum of resonance energies $E_\nu$ can be replaced by integration,
\begin{equation}\label{eq:sum_int}
\sum_\nu\longrightarrow\int \frac{dE_\nu}{D_{J}}.
\end{equation}
Consequently we obtain the statistical capture cross section 
as a sum over the CSF basis states,
\begin{equation}
\label{eq:Jpi-ccs}
\sigma_c= \frac{\pi^2}{k^2}\sum_{k}\frac{(2J_k+1)}{(2J_i+1)}
\Gamma^{(a)}_{k i}
\frac{1}{2\pi}\frac{\Gamma_{\rm spr}}{(\varepsilon-\varepsilon_k)^2+\Gamma_{\rm spr}^2/4}
\end{equation}
where $\varepsilon _k = E_k - E_i$ and
\begin{equation}
\label{eq:Gamma_ki}
\Gamma^{(a)}_{k i}
 = 2\pi \sum_{jl} |\langle (\varepsilon_k jl; J_i)J_k M_k |\hat V |\varphi_k\rangle|^2 \,.
\end{equation}

The total capture strength of any electronic configuration $\tau$
is proportional to the sum over all CSFs $k$ that belong to configuration
$\tau$,
\begin{equation}
\label{eq:StauEnergy}
S_\tau (\eps) = \sum_{k \in \tau} \frac{2J_k+1}{2J_i+1} \Gamma^{(a)}_{k i}
\frac{1}{2\pi}\frac{\Gamma_{\rm spr}}{(\varepsilon-\varepsilon_k)^2+\Gamma_{\rm spr}^2/4}.
\end{equation}
The integral strength
\begin{equation}
\label{eq:Itau}
I_\tau = \int S_\tau(\eps) d\eps 
 = \sum_{k \in \tau} \frac{2J_k+1}{2J_i+1} \Gamma^{(a)}_{k i} \,,
\end{equation}
is determined by the autoionization widths (\ref{eq:Gamma_ki}). Comparing
with Eq.~(\ref{eq:Jpi-ccs}) we see that $S_\tau(\eps)$ gives the contribution of configuration $\tau$ to the reduced capture cross section
$\sigma_c k^2/\pi^2$.

\subsection{Configuration-averaged statistical theory}
\label{sec:CA-theory}

A further approximation may be made as the two-body Coulomb interaction has non-zero value only between the determinants which differ from the target and incident by most two orbitals (the doorways). These configurations may be written in the single-particle basis as $\tau \to \alpha\beta\gamma^{-1}$. We treat  these as the only active electrons, assuming that all other electrons are spectators. We construct CSFs $k$ within $\tau$ and perform the summation over all $J_k$ for a single multiplet within \eref{eq:Jpi-ccs}:
\begin{align}
\label{jsum}
&\sum_{J_k} (2J_k+1)|\langle (\eps jl;\gamma) J_k M_k | V | (\alpha;\beta) J_k M_k \rangle|^2 \nonumber \\
& = \sum_\lambda \frac{X_\lambda[\eps\gamma\alpha\beta]^2}{2\lambda +1} 
 + \sum _{\lambda\lambda '}\left\{ {j_\gamma \atop j }{j_\beta \atop j_\alpha }{\lambda \atop \lambda '}\right\}
    X_\lambda[\eps\gamma\alpha\beta] X_{\lambda'}[\eps\gamma\beta\alpha] \nonumber \\
 &\quad +\alpha\leftrightarrow\beta \,,
\end{align}
where the Coulomb matrix element is
\begin{align}
\label{eq:redmael}
X_\lambda[c\gamma\alpha\beta]
=& (-1)^{\lambda+j_c+j_\gamma+1} \sqrt{[j_c][j_\alpha][j_\gamma][j_\beta]} \\
& \times \xi(l_c +l_\alpha +\lambda )\xi(l_\gamma +l_\beta +\lambda ) \nonumber \\
& \times \left(	{\lambda \atop 0}{j_c \atop -\frac{1}{2}}
		{j_\alpha \atop \frac{1}{2}}\right) \left( {\lambda \atop 0}
		{j_\gamma \atop -\frac{1}{2} }{j_\beta \atop \frac{1}{2}}\right) 
R_\lambda (c\gamma\alpha\beta )\,, \nonumber
\end{align}
$\xi (L)=[1+(-1)^L]/2$ is the parity factor, \hbox{$[j] = 2j +1$}, and
\begin{align*}
R_\lambda (c\gamma\alpha\beta)
 = \iint & \frac{r_<^\lambda} {r_>^{\lambda +1}}[f_c (r)f_\alpha (r)+g_c (r)g_\alpha (r)] \\
 &\times [f_\gamma (r')f_\beta (r')+g_\gamma (r')g_\beta (r')]drdr'
\end{align*}
is the radial Coulomb integral, $f$ and $g$ being the upper and lower components of the relativistic orbital spinors.

The final configuration-averaged capture cross section is obtained as 
\begin{widetext}
\begin{align}
\label{eq:capture-explicit}
 \sigma^\textsl{CA}_c (\eps) = &\frac{\pi^2}{k^2}\sum _{\alpha\beta\gamma}\sum_{jl}
\Biggl[ 
\sum _\lambda \frac{X_\lambda[\eps\gamma\alpha\beta]^2}{2\lambda +1}+\sum _{\lambda\lambda '}\left\{ {j_\gamma 
\atop j }{j_\beta \atop j_\alpha }{\lambda \atop \lambda '}\right\}
X_\lambda[\eps\gamma\alpha\beta] X_{\lambda'}[\eps\gamma\beta\alpha] +\alpha\leftrightarrow\beta
 \Biggr]\nonumber \\
&\times \frac{n_\gamma }{[j_\gamma]}\left(1-\frac{n_\alpha }{[j_\alpha]}\right)\left(1-\frac{n_\beta}{[j_\beta]}\right)
   \frac{\GammaSpr}{(\varepsilon -\varepsilon _{\alpha\beta\gamma^{-1}} )^2 +\GammaSpr^2/4}
\end{align}
\end{widetext}
Here $n_\alpha$, $n_\beta$ and $n_\gamma$ are the occupation numbers of the corresponding subshells (ranging from 0 to $2 j_\alpha+1$, etc). The energy $\eps_{\alpha\beta\gamma^{-1}}$ is the energy for the resonance (relative to the \Wcompound\ ionization threshold). In \cite{dzuba12pra,flambaum02pra} the resonance energy used is the single-particle energy $\eps_{\alpha\beta\gamma^{-1}} = \eps_\alpha + \eps_\beta - \eps_\gamma$, while in this work we present our configuration-averaged statistical theory using the configuration-averaged energy of a relativistic configuration~\cite{gribakin99ajp}
\begin{equation}
\label{eq:CA_energy}
E^\textsl{CA}_i = E_\textrm{core} + \sum_a \epsilon_a n_a + \sum_{a \leq b} \frac{n_a (n_b - \delta_{ab})}{1 + \delta_{ab}} U_{ab}
\end{equation}
where $\epsilon_a$ is the single-particle energy of orbital $a$ in the field of the core, and
$U_{ab}$ is the average of Coulomb matrix elements for the electrons in orbitals $a$ and $b$:
\begin{multline}
\label{eq:Uab}
U_{ab} = \frac{[j_a]}{[j_a]-\delta_{ab}} \Bigg[ R_0 (abab) \\
	-\sum_\lambda \xi(l_a + l_b + \lambda) R_\lambda (abba)
	\left( {j_a \atop \frac{1}{2}} {j_b \atop -\frac{1}{2}} {\lambda \atop 0} \right)^2 \Bigg] . \nonumber
\end{multline}
The number of states in each relativistic configuration is
\begin{equation}
N_i = \prod_a \frac{[j_a]!}{n_a! ([j_a] - n_a)!} \,.
\end{equation}
From \eref{eq:capture-explicit} we also define configuration-averaged capture strengths $S^\textsl{CA}_{\tau} (\eps)$ and $I^\textsl{CA}_{\tau}$ analagously to those of the level-resolved case, i.e. $S^\textsl{CA}_{\tau} (\eps) = \sigma^\textsl{CA}_c k^2/\pi^2$ where the first sum only runs over a single configuration $\tau = \alpha\beta\gamma^{-1}$.

Equation \eref{eq:capture-explicit} is identical to the MBQC theory
formula used in \cite{flambaum02pra,dzuba12pra,dzuba13pra}. The form of Eq.~(\ref {eq:capture-explicit}) is similar to the expressions which emerge in the so-called average-configuration approximation \cite{pindzola86}. The difference between the two approaches is that in a system with chaotic eigenstates, the averaging that leads to \eref {eq:capture-explicit} occurs naturally due to the strong configuration mixing, rather then being introduced by hand to simplify the calculations.
In this work we refer to it as the configuration-averaged (CA) statistical theory, to distinguish it from the level-resolved (\Jpi) calculation \eref{eq:Jpi-ccs} that we will now elucidate.

\section{Calculations}
\label{sec:calc}

For the level-resolved MBQC calculation of Eq.~\eref{eq:Jpi-ccs} we used configuration interaction (CI) to calculate the energies and wavefunctions of \hbox{2\,014\,212} excited levels of \Wcompound, from which we calculate capture widths from the \Wtarget\ ground state. The levels correspond to 63 configurations with configuration-average energies in the range $-13$ to 114\,eV (relative to the \Wcompound\ ionization energy). They are also doorway states for the $\Wtarget + e^-$ recombination process, i.e., they are dielectronic excitations of \Wcompound. All such doorway states have one hole in either the $4d^{10}$ or $4f^8$ shells.

Because diagonalization of the complete Hamiltonian is not practical for this system, each CI Hamiltonian includes all CSFs corresponding to one configuration (which may include many relativistic configurations) and one (\Jpi, $M$) symmetry. While the latter is exact for the Coulomb interaction (which does not mix states of different symmetries), the former constraint must be justified in the context of the statistical theory. In fact, such a choice models precisely the capture cross section into the doorway states (which are not compound resonances), exactly as required by the statistical theory. In the following we denote these levels by the subscript $n$. These are not just the CSFs of Section~\ref{sec:stat-theory}, but rather include CI mixing among all CSFs $k$ corresponding to a configuration. The largest Hamiltonian matrix diagonalised in this work was for the $4d^{-1}\,4f^8\,5p\,5f$ configuration with $J = 9/2$: the number of CSFs (and hence the matrix size) in this case is 25\,112.
An improvement on this method would be to include all dielectronic excitations with a given symmetry into a single CI calculation. Of course even this is computationally difficult, and the overall strength would be unchanged once a sum over all such doorways is performed. The major difference of such a method would be to shift the energies of the levels (within \GammaSpr, and typically by much less than \GammaSpr), but this effect is already modelled by the statistical theory and the change in the continuum electron energy falls within our approximation $\Gamma^{(a)}_{\nu i} \rightarrow \Gamma^{(a)}_{k i}$ (Section~\ref{sec:stat-theory}).

In this work the single-orbital basis functions and the continuum wavefunctions are both calculated in the $V^N$ potential of the target ion. The target orbitals are constructed by solving the relativistic Hartree-Fock equations for the configuration [Kr] $4d^{10}\,4f^8$. The open $4f$-shell is treated by scaling the corresponding interactions for the closed shell by the factor $\nicefrac{8}{14}$. We then use $B$-splines~\cite{johnson88pra} to calculate excited states up to $8spdfg$; the results obtained with this basis are very close to those of excited Hartree-Fock orbitals. We found that the electron capture strengths are saturated when continuum orbitals with $l \leq 6$ are included. Finally, to speed up our calculations of Eq.~\eref{eq:Gamma_ki} we used a grid of continuum functions separated at 10\,eV intervals and starting at 0.1\,eV above the ionization threshold. Thus, rather than calculating $\Gamma^{(a)}_{n i}$ with continuum energy $\eps_n$, we actually calculate $|\langle (\eps_{\rm grid} jl; J_i)J_k M_k |\hat V |\varphi_k\rangle|^2$ with $\eps_{\rm grid}$ being the grid point closest to $\eps_n$. The associated error is negligible, since the matrix elements depend weakly on the energy of the wavefunctions in the ionic Coulomb field, when normalized to the $\delta $-function of energy.

Our configuration-averaged cross sections are obtained from Eq.~(\ref{eq:capture-explicit}) using the same single-particle orbitals as the level-resolved calculation. We choose occupation numbers of the target ground state based on a CI calculation of the target $4f^8$ manifold with $J = 6$, which gives relativistic occupations $4f_{5/2}^x\,4f_{7/2}^{8-x}$ with $x = 4.62714$. From these we generate the occupation numbers for the $4f$ orbitals of the doorways in Eqs.~\eref{eq:capture-explicit} and \eref{eq:CA_energy}, subtracting one from the relevant orbital depending on where the hole is.

\section{Results and Discussion}
\label{sec:results}

In this section we examine the results of our level-resolved calculations described in Section~\ref{sec:calc} and compare them with results obtained from the configuration-averaged MBQC statistical theory (Section~\ref{sec:CA-theory}) performed with the same single-particle orbital basis.

\subsection{Effect of conformation of the $4f$ `spectator' electrons}
\label{sec:conformation}

\begin{figure}[tb]
\centering
\includegraphics[width=0.45\textwidth]{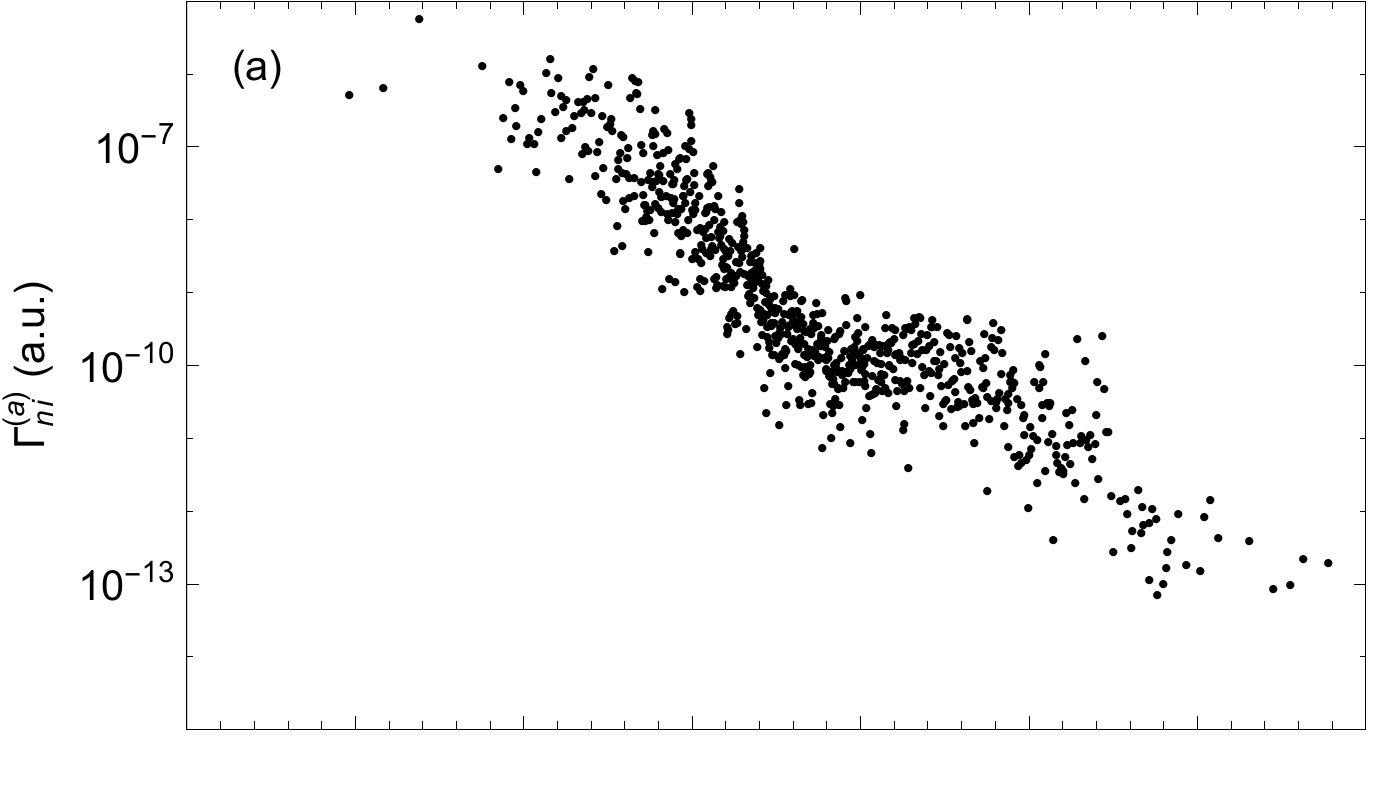}
\includegraphics[width=0.45\textwidth]{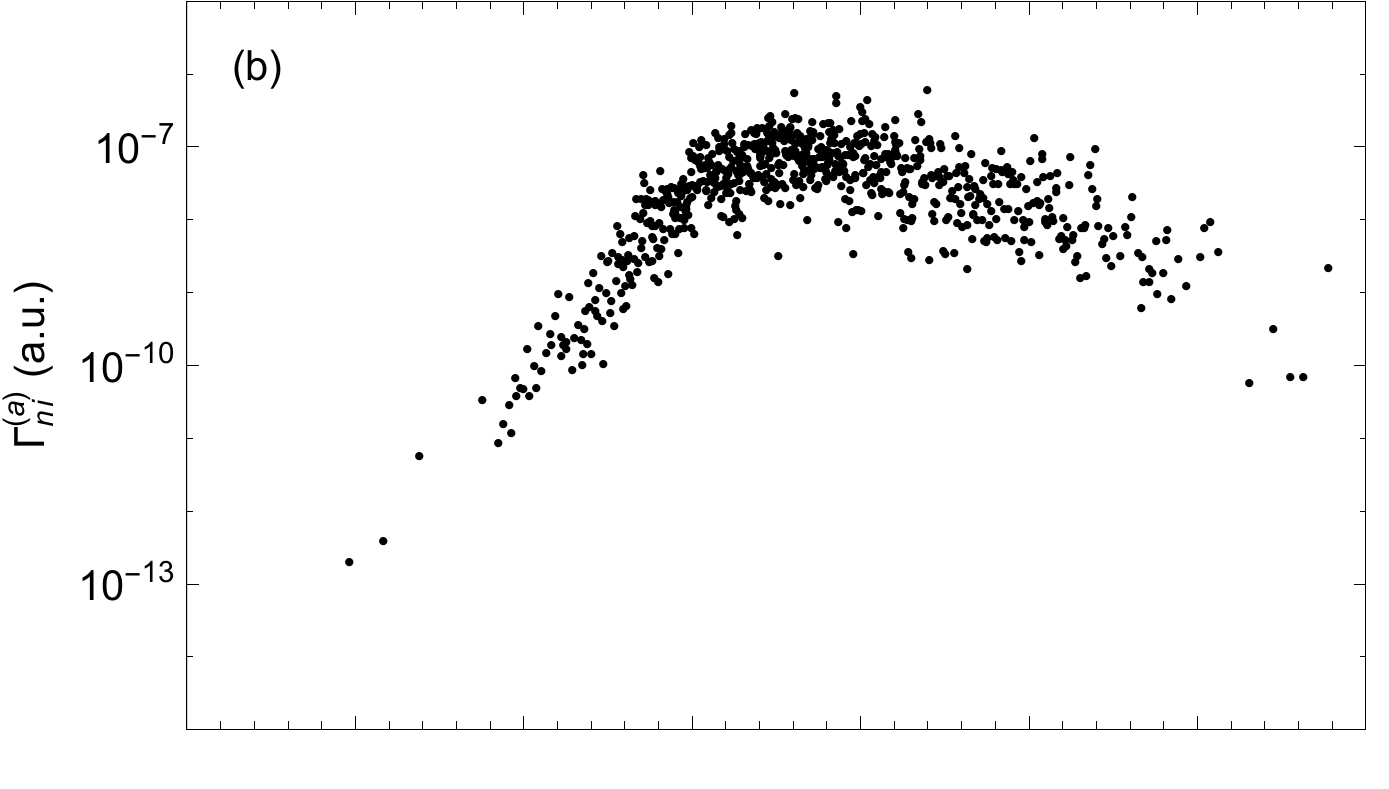}
\includegraphics[width=0.45\textwidth]{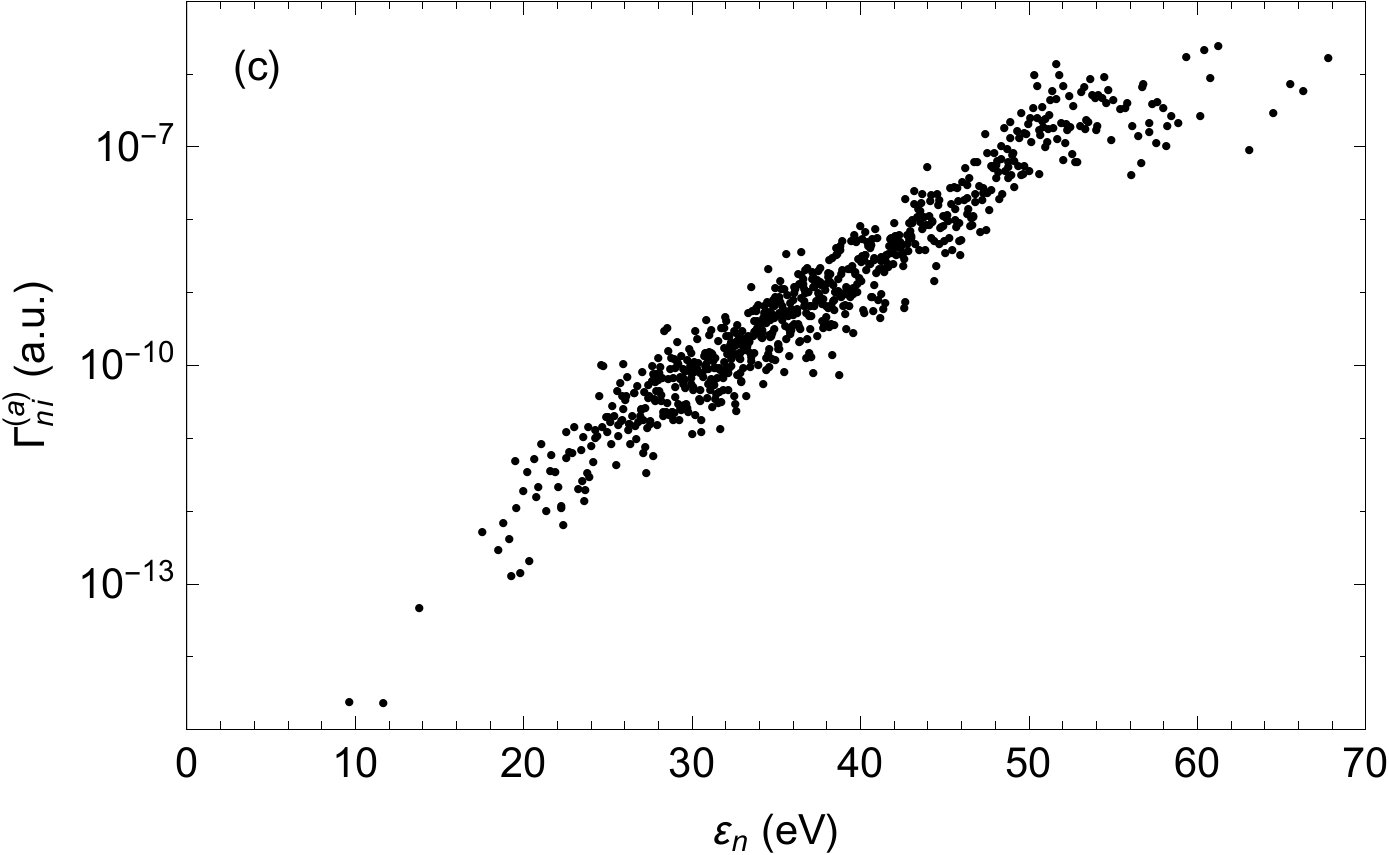}
\caption{\label{fig:targetlevel} Autoionization widths $\Gamma^{(a)}_{\nu i}$ for $J = 11/2$ levels of the configuration $4f^7 6s\,6d$.
(a) Capture from the ground state $4f^8\ J = 6$;
(b) Capture from the 18\textsuperscript{th} excited state in the $4f^8\ J = 6$ manifold;
(c) Capture from the 37\textsuperscript{th} (highest) excited state in the $4f^8\ J = 6$ manifold.}
\end{figure}

One immediately noticeable effect in the level-resolved calculation is that levels with lower energies within a single configuration tend to have larger capture widths. A typical example is seen in \Fig{fig:targetlevel}(a), which shows the strengths of all 780 levels of the $4f^7 6s\,6d$ configuration with $J = 11/2$. Note the logarithmic scale on the vertical axis: different levels of the same configuration with the same $J$ have widths that differ by up to 8 orders of magnitude. 

This trend is apparently due to the effect of the $4f^7$ core which, rather than being a `spherical' potential as in the configuration-averaged model, actually has a large number of possible ``conformations''. Levels where the $4f^7$ spectators are coupled in a way most similar to those of the target, maximise the capture strength, and these same conformations tend to have lower energy by virtue of the fact that the target ground state is a low-energy conformation. We tested this  understanding by calculating the autonionization widths for different states of the $4f^8$ configuration with $J = 6$ (there are 38 such levels). The results in \Fig{fig:targetlevel}(c) correspond to the target being in the highest-energy $4f^8$ conformation, and the trend of \Fig{fig:targetlevel}(a) is entirely reversed: the highest energy levels of the $4f^7 6s\,6d\ (J = 6)$ configuration now have the largest strengths. \Fig{fig:targetlevel}(b) shows the strengths of capture from the 18\textsuperscript{th} excited level of the same manifold, which favours conformations with energies around the middle of the possible range. Thus we see that electron coupling within the $4f^7$ core has a very large effect on the capture cross section.

\subsection{Effect of angular momentum}
\label{sec:angular}

\begin{figure}[tb]
\centering
\includegraphics[width=0.45\textwidth]{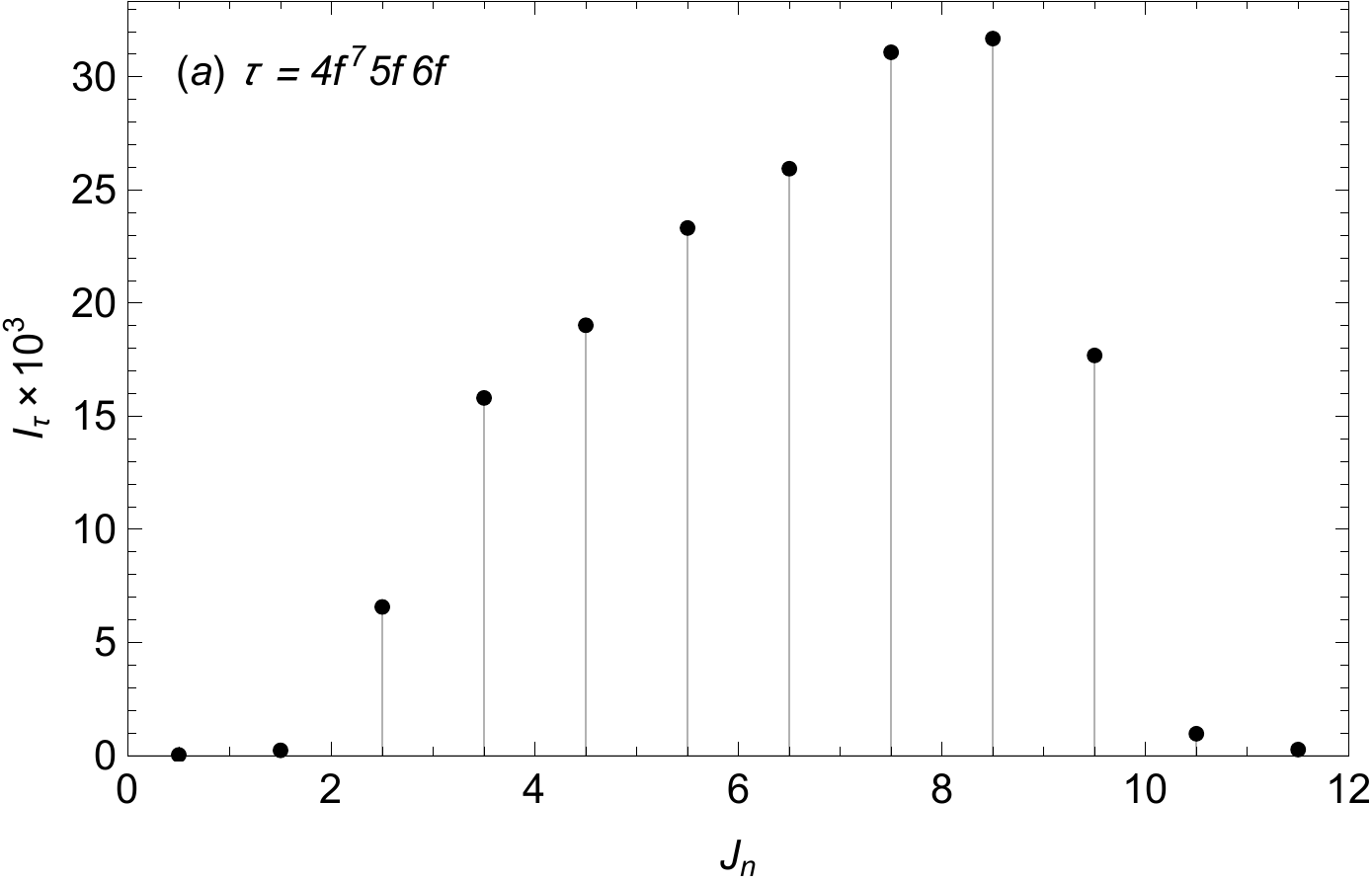}
\includegraphics[width=0.45\textwidth]{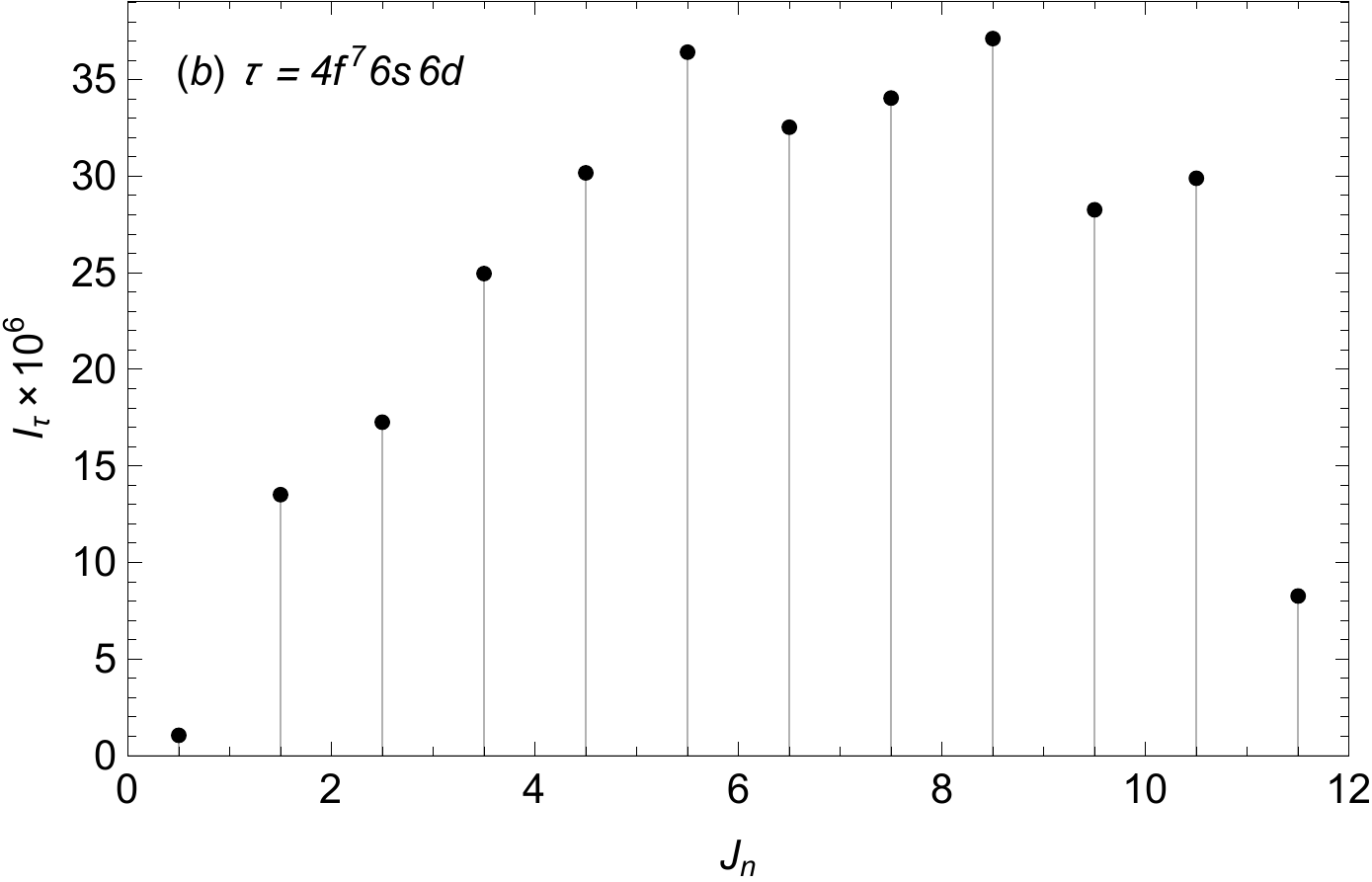}
\caption{\label{fig:Jeffect}
Contributions of states with different angular momenta $J_n$ to the capture
strength Eq.~\eref{eq:Itau} for different configurations $\tau$:
(a) $\tau = 4f^7 5f\,6f$; (b) $\tau = 4f^7 6s\,6d$.
}
\end{figure}

In our level-resolved calculation we limit the orbital angular momentum of the continuum (incoming) electron to $l \leq 6$ (i.e., $j \leq 13/2$). Since the \Wtarget\ target ground state has $J = 6$, only resonances with $1/2 \leq J_n \leq 25/2$ contribute. The different values of $J_n$ contribute in different ways to the integrated strength of a configuration $I_\tau$. In \Fig{fig:Jeffect} we show these contributions for two different configurations with vastly different integral strengths. The configuration $4f^7 5f\,6f$ is one of the largest contributors to the total capture cross section in the range 0--100\,eV, while the configuration $4f^7 6s\,6d$ is much weaker. They show rather different trends with $J_n$, but generally we observe no strong dependence on the total angular momentum, though usually the resonances with small or large $J_n$ do not contribute as much as those nearer to $J=6$.

\subsection{Comparison of level-resolved and configuration-averaged statistical theory}

\begin{figure*}[htb]
\centering
\includegraphics[width=\textwidth]{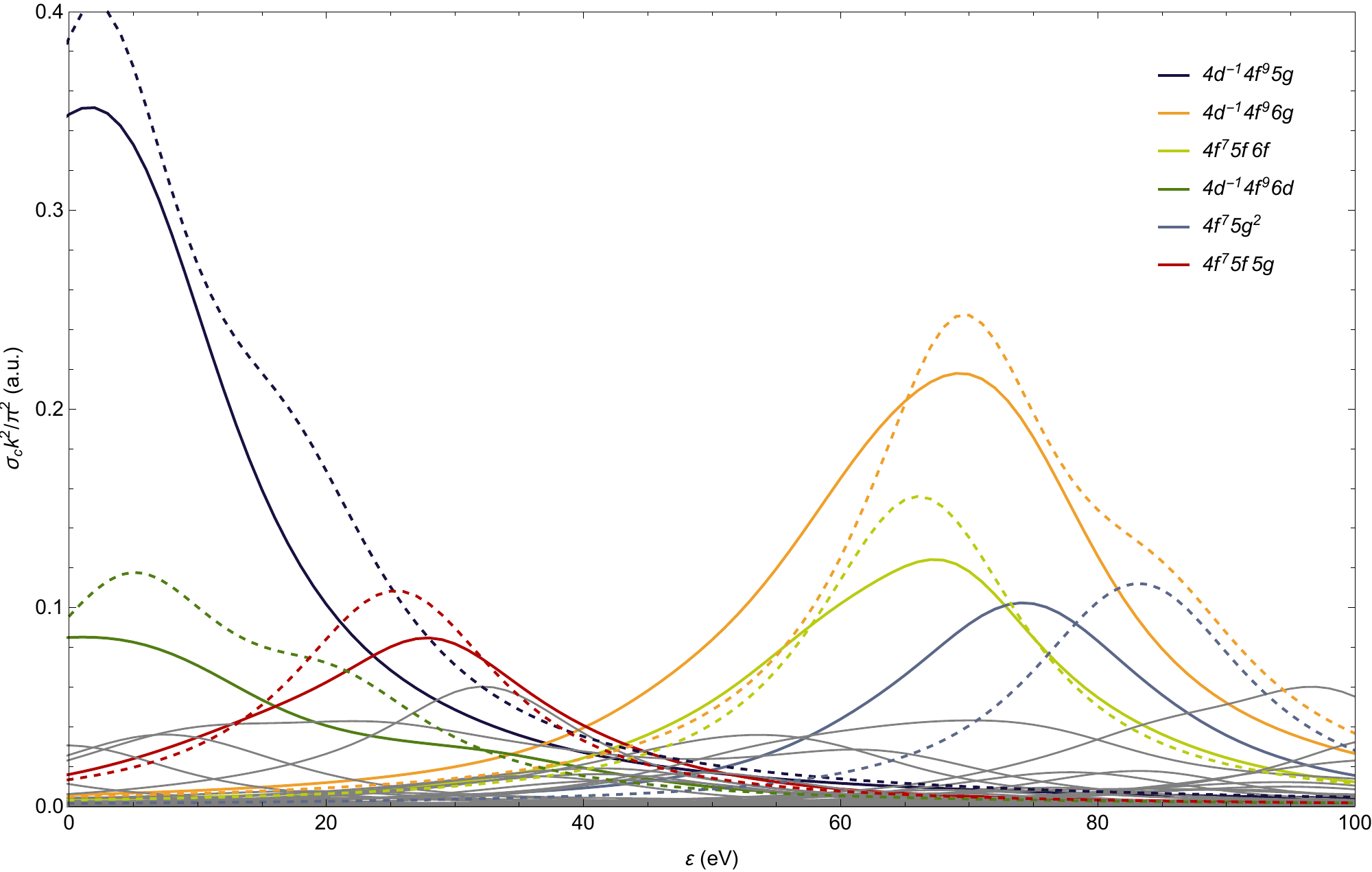}
\caption{\label{fig:CAcomparison} (color online)
Comparison of the level-resolved capture strength $S_\tau (\eps)$ [Eq.~(\ref{eq:StauEnergy}), solid lines] and configuration-averaged 
$S^\textsl{CA}_\tau (\eps)$ (dashed lines) for six configurations $\tau$ which give the largest contribution to the capture cross section.
Capture strengths of other configurations are shown in grey (level-resolved calculation only). We take $\GammaSpr = 0.68\,\textrm{a.u.} = 18.5$\,eV~\cite{dzuba12pra}.
}
\end{figure*}

The capture strengths and doorway-state energies of the level-resolved theory can be directly compared with the configuration-averaged theory by examining
the capture strengths $S_\tau (\eps)$ \eref{eq:StauEnergy} of individual configurations from both calculations.
The capture cross section is dominated by a few configurations, and we show the comparison between the two theories for these configurations in \Fig{fig:CAcomparison}. All other configurations are shown as grey lines and we have not presented $S^\textsl{CA}_\tau (\eps)$ for these. Note that we take $\GammaSpr = 0.68\,\textrm{a.u.} = 18.5$\,eV calculated in~\cite{dzuba12pra}.

\Fig{fig:CAcomparison} shows good agreement between the two statistical theories. That is, despite the strong effect of the conformation of the spectator electrons and the complexity introduced by angular momentum, when summed over all levels the two theories are in remarkable agreement. In our discussion of the effect of the spectator electrons in Section~\ref{sec:conformation} we saw that levels with lower energies have larger capture widths. This suggests that in the level-resolved theory the peaks should be shifted to lower energies relative to the CA MBQC calculation, since the greatest contribution is given by doorway states with lower energies. Indeed, this is seen in \Fig{fig:CAcomparison}, although the difference is perhaps not as large as one might expect when considering the spread of level energies (see \Fig{fig:targetlevel}). The reason for this is that part of the  conformation effect comes from the relativistic occupancy of the target ground state. Our CI calculation of the target gives average occupancies $4f_{5/2}^{4.6}\,4f_{7/2}^{3.4}$ (Sec.~\ref{sec:calc}), and these are the occupancies $n_\gamma$ used in \eref{eq:capture-explicit} and \eref{eq:CA_energy}. Thus the configuration-average calculation itself is shifted to lower energies relative to the average energy of the resonance levels, reducing the difference between the CA and level-resolved calculations.

The total height of the peaks is generally smaller in the case of the level-resolved (\Jpi) theory. However, we find that the total (integrated) strength $I_\tau$ always agrees with the CA theory to within around 15\%. Rather, the level resolution in the \Jpi\ theory spreads the capture strength over a wider range of energy, broadening the peak while maintaining the integrated strength. Thus while any individual configuration appears to be broader and lower in the \Jpi\ theory, when summed over all configurations, the total capture cross section is almost unchanged.

\begin{figure}[tb]
\centering
\includegraphics[width=0.45\textwidth]{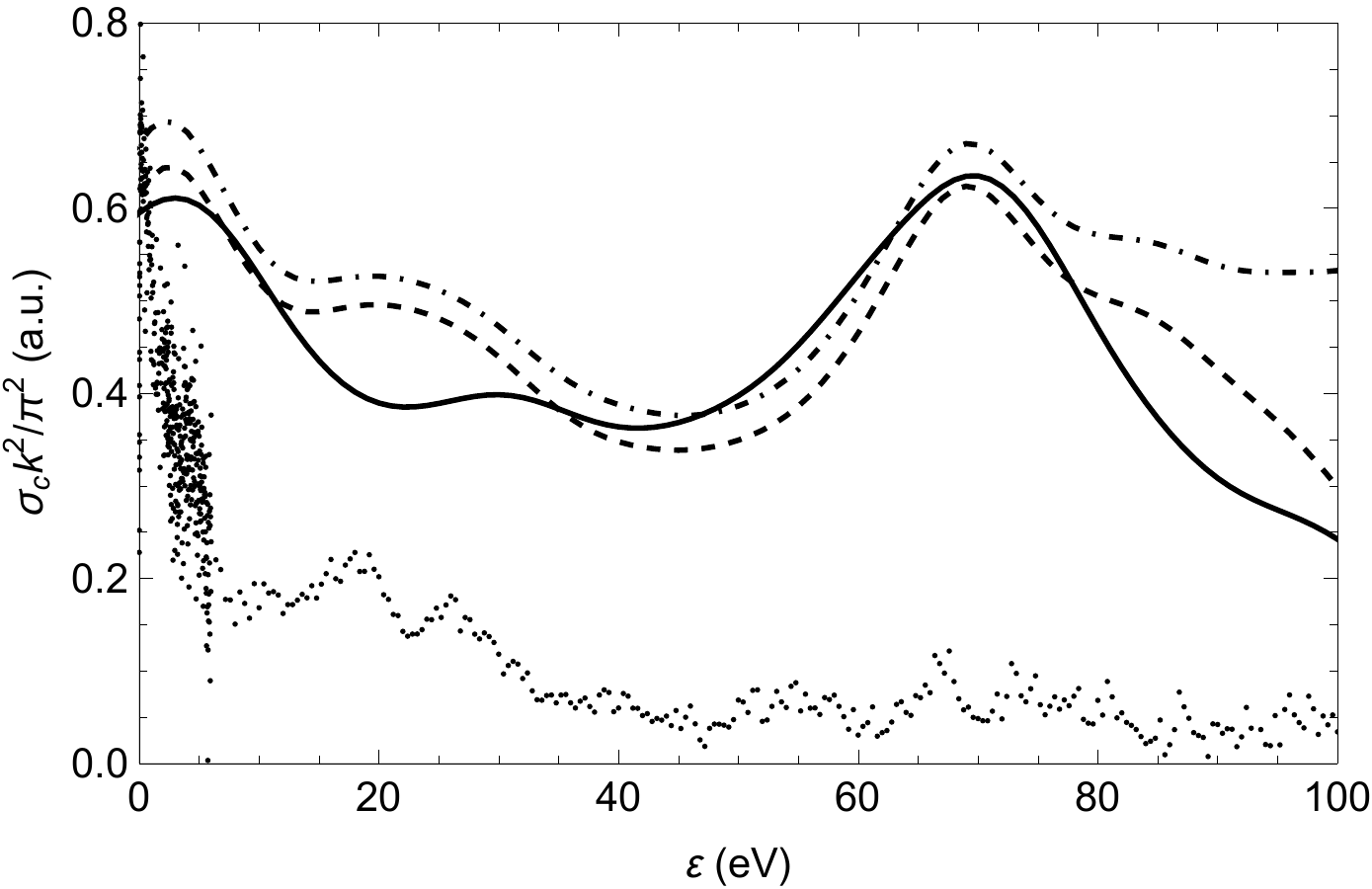}
\caption{\label{fig:total} Total reduced capture cross section in the level-resolved calculation (solid) and the configuration-averaged calculation (dashed). The dot-dashed curve is a configuration-average calculation including configurations from a wider range of energy, from $-100$ to 200\,eV. The dots show the unaveraged experimental recombination data of \cite{schippers11pra}.}
\end{figure}

The total reduced capture cross section $\sigma_c k^2/\pi^2$ in the different calculations is presented in \Fig{fig:total}. As would be expected from the preceding discussion, we find good agreement between our two theories over the entire range from threshold to 100\,eV (the solid line is level-resolved and the dashed line is the CA MBQC calculation). As explained in Sec.~\ref{sec:calc}, we have only included configurations where the CA energy lies between $-13$ and 113\,eV. However, the Lorentz distribution is heavy-tailed, and with $\GammaSpr=18.5$\,eV configurations from outside this range could have an effect if they possess particularly strong capture cross sections. Therefore in \Fig{fig:total} we include another CA calculation (shown as the dot-dashed curve) where configurations with energies in the range $-100$ to 200\,eV are included. We see that indeed, there is some contribution from configurations outside the range of our main calculations. In \Fig{fig:total} we also include the experimental recombination data of \cite{schippers11pra}. Since in this paper we have neglected the fluorescence yield, a direct comparison cannot be made except very close to threshold, $\eps \lesssim 1$\,eV, where the radiative yield should be very close to unity. We see that the calculated capture cross section is still slightly below the experimental recombination rate at very low energy. We note, however, that the statistical theory cannot resolve individual sharp resonances, which may occur close to threshold.

\begin{figure}[htb]
\centering
\includegraphics[width=0.45\textwidth]{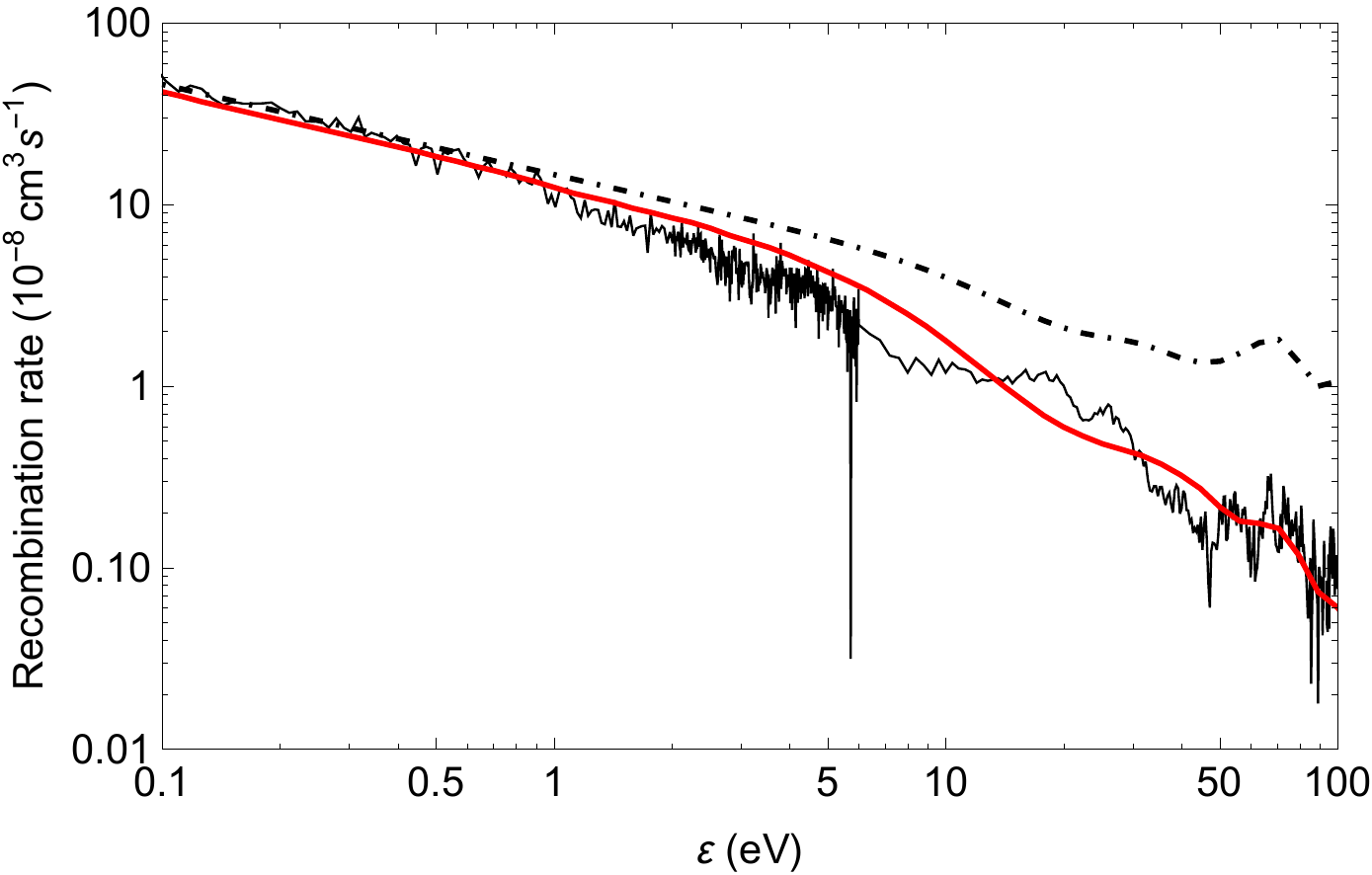}
\caption{\label{fig:recombination} (color online) Recombination rate from experiment~\cite{schippers11pra} (black) and the theoretical results of this paper (red). The latter is obtained by multiplying our capture rate (dot-dashed line) with our extracted fluorescence yield~\eref{eq:omegafit}.}
\end{figure}

Using the data in \Fig{fig:total} we can extract the fluorescence yield $\omega_f$ (see Eq.~\ref{bratio}). As shown in~\cite{dzuba12pra}, the fluorescence yield can be approximated as
\[
\omega_f (\eps) \approx \frac{1}{1+a N(\eps)}
\]
where $N(\eps)$ counts the number of autoionization channels open at an energy $\eps$, that is, the number of \Wtarget\ target states with energy below \eps. This form mimics the increase of the autoionization width of the resonances vs their radiative width. We have calculated $N(\eps)$ over our range of interest using the same kind of level-resolved calculation outlined in Sec.~\ref{sec:calc}, and find that using $a = 0.012$ leads to a good fit to experimental data over our energy range (0 to 100\,eV). In fact, we observe that over this energy range $N(\eps) \propto \eps$ and find that $\omega_f$ can be well approximated by using the considerably simpler function
\begin{equation}
\label{eq:omegafit}
\omega_f (\eps) = \frac{1}{1 + b\,\eps}
\end{equation}
with $b = 0.124$~eV$^{-1}$. Multiplying our final capture cross section by this factor leads to our final radiative recombination rate, which is compared with experiment in \Fig{fig:recombination}.

\section{Conclusion}

We have derived a level-resolved many-body quantum chaos statistical theory of resonant electron capture in highly charged ions that includes the effects of spectator electrons and respects the total angular momentum $J$ of the resonances. We have calculated the energies and wavefunctions of 2.0\E{6} doubly excited doorway states of \Wcompound\ in this level-resolved theory, and have determined the capture cross section by the ground-state \Wtarget\ target for each of these. We found rather strong dependence on the conformation of the spectator electrons and  a relatively weak dependence on the angular momentum of the individual resonances. Nevertheless, we observe that when considering the sum over all resonances, the configuration-average MBQC statistical theory used in previous works \cite{flambaum02pra,dzuba13pra,dzuba12pra} is robust with respect to inclusion of these effects.

A reasonable course of calculation in the future would be to first determine the most important configurations using the configuration-average formulation of the statistical theory, and then perform a more accurate level-resolved calculation on those that contribute the most to the cross section.
Not calculated in this work is the fluorescence yield, $\omega_f$ \eref{bratio}, which has a large influence on the recombination rate at energies that are not very close to threshold. In this case one might predict a stronger effect from the angular momentum quantum numbers due to the selection rules in play.
In the future our level-resolved statistical theory can also be applied to other processes where chaotic mixing plays an important role, such as photo- and electron-impact ionization and scattering processes in highly charged ions~\cite{flambaum15pra}.

\begin{acknowledgments}
This work is supported by the Australian Research Council.
GG is grateful to the Faculty of Science (UNSW) for supporting his visit.
\end{acknowledgments}

\appendix*
\section{Theory of resonant recombination}
The following derivation of the resonant recombination cross section is based on the formulation of Ref.~\cite{gribakin05pra}.
The initial state (${\bf p}$, $\mu$) describes the electron with momentum ${\bf p}$ and helicity $\mu={\bf \sigma}\cdot{\bf p}/2p=\pm 1/2$ incident on the target ion W$^{20+}$ in the ground state $|J_iM_i\rangle$.
Let us expand the incident electron state in partial waves,
\begin{equation}\label{free_wf}
|{\bf p}, \mu\rangle =\frac{(2\pi)^{3/2}}{\sqrt{p}}
\sum_{jlm}\langle\Omega_{jlm}(\hat {\bf p})|\chi_\mu(\hat {\bf  p})\rangle i^le^{i\delta_{jl}}|\varepsilon jlm\rangle,
\end{equation}
where $\Omega_{jlm}$ and $\chi_\mu$ are spherical and ordinary spinors and $\delta_{jl}$ is the scattering phase shift. The wave function (\ref{free_wf}) is normalized so that $\langle {\bf p}',\mu'|{\bf p},\mu\rangle=(2\pi)^3\delta({\bf p}'-{\bf p})\delta_{\mu'\mu} $ and the radial functions are normalized to the delta function of energy, $\langle \varepsilon'j'l'm'| \varepsilon j l m \rangle=\delta( \varepsilon'- \varepsilon )\delta_{j'j} \delta_{l'l}\delta_{m'm}$. The spinor matrix element in (\ref{free_wf}) is
\begin{align}\label{eq:spinor_mat}
\langle \Omega_{jlm}(\hat {\bf p})|\chi_\mu (\hat {\bf  p})\rangle  =& \sum_\lambda C^{jm}_{l\lambda \frac{1}{2} \mu} Y_{l\lambda}^\ast(\hat {\bf  p}),
\end{align}
where $C^{jm}_{l\lambda \frac{1}{2} \mu}$ is the Clebsh-Gordon coefficient and $Y_{l\lambda}$ is the spherical harmonic.

In the independent-process approximation, the two paths (direct radiative or resonant) for recombination are summed incoherently.
The amplitude of resonant recombination is
\begin{equation}\label{amplitude}
A  = \sum_{\nu} \frac{i\sqrt{2\pi\omega/V}\langle n |{\bf e}_q \cdot {\bf D} |\nu\rangle\langle \nu |\hat V |{\bf p}, \mu ; J_iM_i\rangle}
{E_i+\varepsilon-E_\nu+i\Gamma_\nu/2},
\end{equation}
where ${\bf e}_q$ and $\omega$ define the polarisation and frequency of the photon, and $\hat V$ is the Coulomb interaction.The dipole approximation ${\bf D}=-\sum_{j} e {\bf r}_j$ is used for radiative transition. $V$ is the quantisation volume for the electromagnetic field, and $\Gamma_\nu$ is the total width of the resonance state $|\nu\rangle$.
The corresponding cross section is
\begin{equation}\label{cs0}
\sigma = \frac{2\pi}{p} \sum_{q,n}\int |A|^2 \delta(E_i+\varepsilon -\omega -E_n) V \frac{\omega^2 d\omega d\Omega}
{(2\pi c)^3}
\end{equation}
In the isolated-resonance approximation, it can be written as follows after integration over $\omega$:
\begin{equation}\label{cs1}
\sigma= \frac{1}{p} \sum_{q,n}\sum_{\nu}\int d\Omega\frac{\omega^3_{\nu n}}{2\pi c^3}\frac{|{\bf e}_q\cdot\langle n |{\bf D} |\nu\rangle|^2|\langle \nu |\hat V |{\bf p}, \mu ; J_iM_i\rangle|^2}
{(E_i+\varepsilon-E_\nu)^2+\Gamma_\nu^2/4}
\end{equation}
Using the definition of autoionization width of the resonance $|\nu\rangle$
\begin{align}\label{auto_width0}
\Gamma_{\nu i}^{(a)}&=\frac{p}{\pi(2J_\nu+1)}\sum_{M_\nu M_{i}\mu} \int \frac{d\Omega}{4\pi} |\langle {\bf p}, \mu ; J_{i}M_{i} |\hat V |\nu\rangle|^2
\end{align}
and the radiative width for transition to the final state $|n\rangle$
\begin{align}\label{rad_wid0}
\Gamma_{\nu n}^{(r)}&=\frac{1}{(2J_\nu+1)}\sum_{q M_n M_\nu} \int d\Omega \frac{\omega^3_{\nu n}}{2\pi c^3} |{\bf e}_q\cdot\langle n |{\bf D} |\nu\rangle|^2 , 
\end{align}
the final expression of the cross section is obtained as
\begin{equation}\label{cs2}
\sigma= \frac{\pi}{p^2}\sum_{\nu}\frac{(2J_\nu+1)}{2(2J_i+1)}\frac{\Gamma_{\nu i}^{(a)} \Gamma_{\nu}^{(r)}}
{(\varepsilon-\varepsilon_\nu)^2+\Gamma_\nu^2/4},
\end{equation}
where $\varepsilon_\nu=E_\nu-E_i$ is the energy of the resonance with respect to the threshold,  and $\Gamma_\nu^{(r)}=\sum_n\Gamma_{\nu n}^{(r)} $  is the total radiative width of the resonance. 

Using the orthogonality relations of spherical harmonics, 
$$\int d\Omega\,Y_{l\lambda}(\hat {\bf  p})Y^\ast_{l'\lambda'}(\hat {\bf  p})=\delta_{ll'}\delta_{\lambda\lambda'} $$ and of Clebsh-Gordon coefficients, $$ \sum_{\mu\lambda} C^{jm}_{l\lambda \frac{1}{2} \mu}C^{j'm'}_{l\lambda \frac{1}{2} \mu}=\delta_{jj'}\delta_{mm'} $$
we obtain the autoionization formula
\begin{widetext}
\begin{align}\label{eq:aut_wid}
\Gamma_{\nu i}^{(a)}&=\frac{2\pi}{2J_\nu+1}\sum_{M_\nu M_{i}\mu} \int d\Omega \left |\sum_{jlm}\sum_{\lambda}C^{jm}_{l\lambda \frac{1}{2} \mu} Y_{l\lambda}(\hat {\bf  p})(-i)^l e^{-i\delta_{jl}}\langle \varepsilon_\nu jlm; J_{i}M_{i} |\hat V |\nu \rangle \right |^2\nonumber \\
&=\frac{2\pi}{2J_\nu+1}\sum_{M_\nu M_{i}}\sum_{jlm}\left |\langle \varepsilon_\nu jlm; J_{i}M_{i} |\hat V |\nu \rangle \right |^2.
\end{align}
\end{widetext}
A further simplification is possible by coupling the continuum orbital $| \eps jlm\rangle$ with the target ground state $|J_{i}M_{i}\rangle$ to construct the constant angular momentum state $|J,M\rangle$:
\begin{align}
 | \varepsilon jlm; J_{i}M_{i} \rangle=\sum_{JM}C^{JM}_{jmJ_iM_i}| (\varepsilon jl; J_{i})JM \rangle.
\end{align}
Substituting it in (\ref{eq:aut_wid}) provides the compact formula for autoionization width:
\begin{equation}\label{eq:aut_wid1}
\Gamma_{\nu i}^{(a)}=2\pi\sum_{jl}|\langle (\varepsilon_\nu jl; J_i)J_\nu M_\nu |\hat V |\nu\rangle|^2
\end{equation}
The radiative width (\ref{rad_wid0}) is also simplified by standard manipulations to  
\begin{align}\label{rad_wid1}
\Gamma_{\nu n}^{(r)}&=\frac{1}{2J_\nu+1}\sum_{M_\nu M_n} \frac{4}{3}\left( \frac{\omega_{\nu n}}{c}\right)^3 |\langle J_n M_n|{\bf D}|J_\nu M_\nu\rangle|^2\nonumber\\
&=\frac{1}{2J_\nu+1}\frac{4}{3}\left( \frac{\omega_{\nu n}}{c}\right )^3 |\langle J_n  \| D \| J_\nu \rangle|^2
\end{align}
where $\omega_{\nu n}=E_\nu-E_n >0$ and $\langle J_n  \| D \| J_\nu \rangle $ is the reduced matrix element of the dipole transition.


\end{document}